\documentstyle[12pt]{article}
\begin{document}
\begin{center}
{\bf{A NEW INTERPRETATION OF THE SEIBERG WITTEN MAP}}
\vskip 2cm
Subir Ghosh\\
\vskip 1cm
Physics and Applied Mathematics Unit,\\
Indian Statistical Institute,\\
203 B. T. Road, Calcutta 700108, \\
India.
\end{center}
\vskip 3cm
{\bf Abstract:}\\
In an alternative interpretation, the Seiberg-Witten map is shown
to be induced by a field  dependent co-ordinate transformation
connecting noncommutative and ordinary space-times. Furthermore, following our previous ideas, it
has been demonstrated here that the above (field  dependent co-ordinate) transformation can occur naturally in the Batalin-Tyutin extended space version of the
relativistic spinning particle model, (in a particular gauge).
There is no need to postulate the space-time non-commutativity in
an {\it ad hoc} way: It emerges from the spin degrees of freedom.

\newpage
As a natural generalization of the phase space Non-Commutativity (NC) in quantum mechanics, NC in {\it spacetime} was  originally introduced by Snyder \cite{sn} as a regularization to tame the short distance singularities, inherent in a Quantum Field Theory (QFT). This is because NC in spacetime can introduce a lower bound in the continuity of spacetime, just as $\hbar$ does in the phase space in quantum mechanics. The advantage of NC as a regularization is that the computational scheme requires very little changes from the ordinary spacetime and in some forms of NC \cite{sn}, (for more recent works see \cite{lor, sg3,mor}), manifest Lorentz invariance can be maintained. However, due to the advent of renormalization techniques in QFT, the idea of Snyder \cite{sn} did not gain much popularity. Also now we know \cite{rev} that the NC prescription does not quite render a QFT well-defined, as it was envisaged somewhat naively.

In more recent times, existence of non-commutativity in  (open) string boundaries in the
presence of a constant two-form Neveu-Schwarz field, and the
resulting Non-Commutative Quantum Field Theory (NCQFT) in the
branes to which the open string endpoints are attached, has rekindled the
interest \cite{rev} in physics in non-commutative space-time. Seiberg and Witten \cite{sw} have shown that the appearence of NCQFT is dependent on the choice of regularization and in fact a QFT in ordinary spacetime and an NCQFT both can describe the same underlying QFT. Concretization of this idea has lead to the celebrated Seiberg-Witten Map (SWM)
\cite{sw} which plays a pivotal role in our understanding of the NCQFT
by directly making contact between NCQFT and QFT in ordinary
space-time via the SWM. At least to the lowest non-trivial order in
$\theta _{\mu\nu} $, the non-commutativity parameter, 
\begin{equation}
[x_{\mu},x_{\nu}]=i\theta_{\mu\nu},
\label{0}
\end{equation}
the SWM can be
exploited to convert a NCQFT to its counterpart living in ordinary space-time,
in which the effects of non-commutativity appears as
local interaction terms, supplemented by $\theta _{\mu\nu} $. In the more popular form of NCQFT, $\theta _{\mu\nu} $ is taken to be constant. This can lead to very striking signatures in particle physics  phenomenology  in the form of Lorentz symmetry breakdown, new interaction vertices etc. \cite
{phen}.

However, as it stands, the SWM  is linked exclusively to NC {\it gauge}
theory, since the original derivation of the SWM \cite{sw} hinges
on the concept of identifying gauge orbits in NC and ordinary
space-times. In the explicit form of the SWM \cite{sw}, the
non-commutativity of the {\it space-time} in which the NC gauge
field lives, is not manifest at all since the map is a relation
between the NC and ordinary gauge fields and gauge transformation
parameters, all having ordinary space-time coordinates as their
arguments.

On the other hand, possibly it would have been more
natural to consider first a map between NC and ordinary space-time
and subsequently to induce the SWM through the  change in the
space-time argument of the gauge field from the ordinary to NC
one. Precisely this type of a geometrical reformulation of the SWM
is presented in this  paper.

In the canonical quantization prescription, the Poisson Bracket algebra is elevated to quantum commutator algebra by the replacement 
$$\{A,B\}\rightarrow \frac{1}{i}[\hat A,\hat B].$$ But presence of {\it constraints} may demand a modification in the Poisson Bracket algebra, leading to the Dirac Bracket algebra \cite{d}, which are subsequently identified to the commutators,
$$\{A,B\}_{DB}\rightarrow \frac{1}{i}[\hat A,\hat B].$$ However, complications in this formalism can arise, (in particular in case of non-linear constraints), where the Dirac Bracket algebra itself becomes operator valued. To overcome this, Batalin and Tyutin \cite{bt} have developed a systematic scheme in which all the physical variables are mapped in an extended canonical  phase space, consisting of auxiliary degrees of freedom besides the physical ones, with all of them enjoying canonical free Poisson Bracket algebra. In this formalism, the ambiguity of using (operator valued) Dirac Brackets as quantum commutators does not arise.

In the spinning particle model \cite{sg1} the canonical $\{x_{\mu},x_{\nu}\}=0$ Poisson Bracket changes to an operator valued Dirac Bracket,
\begin{equation}
\{x_{\mu},x_{\nu}\}_{DB}=-\frac{S_{\mu\nu}}{M^{2}},
\label{dir}
\end{equation}
due to the presence of constraints. In the above, the dynamical variable $S_{\mu\nu}$ represents the spin angular momentum and $M$ is the mass of the particle. This forces us to exploit the Batalin-Tyutin prescription \cite{bt}.

In a recent paper \cite{sg3}, we have constructed a mapping of the form,
\begin{equation}
\{x_\mu ,x_\nu \}=0~~,~~
x_\mu \rightarrow \hat x_\mu ~~;~~
\{\hat x_\mu ,\hat x_\nu \}=\theta _{\mu\nu},
\label{nc}
\end{equation}
which bridges the noncommutative and ordinary space-times. Note that $\hat x_{\mu}$ lives in the Batalin-Tyutin \cite{bt} extended space and is of the generic form $\hat x_{\mu}=x_{\mu}+X_{\mu}$, where $X_{\mu}$ consists of physical and auxiliary degrees of freedom. Explicit expressions for $X_{\mu}$ are to be found later \cite{sg3}.

This space-time map induces in a natural way the following map between noncommutative and ordinary degrees of freedom,
\begin{equation}
\lambda (x)\rightarrow \lambda (\hat x)
\rightarrow \hat \lambda (x)~~,~~A_\mu (x)\rightarrow A_\mu (\hat x)
\rightarrow \hat A_\mu (x).
\label{2}\end{equation}
Here $\hat \lambda$ and $\hat A_\mu$ are the NC counterparts of $\lambda$ and $A_\mu$, the
abelian gauge transformation parameter and the gauge field respectively and $\hat x_\mu $
and $x_\mu $ are the NC and ordinary space-time co-ordinates.

On the other hand, there also exists the SWM \cite{sw} which interpolates between noncommutative and ordinary variables,
\begin{equation}
\lambda (x)
\rightarrow  \hat \lambda (x)~~,~~A_\mu (x)\
\rightarrow \hat A_\mu (x).
\label{2a}\end{equation}

It is only logical that the above two schemes ((\ref{nc}-\ref{2}) and \ref{2a}) can be related. In the present work we have precisely done that. The formulation \cite{sg3}
(\ref{nc}-\ref{2}) being the more general one, we have explicitly demonstrated how it can be reduced to the SWM \cite{sw}, in a particular gauge. This incidentally demonstrates the correctness of the procedure. The above idea was hinted in \cite{sg3}.
\footnote {The present analysis being classical, (non)commutativity is to be interpreted
in the sense of Poisson or Dirac Brackets.}

In this context, let us put the present work in its proper perspective. Recently a
number of works \cite{corn} have appeared with the motivation of recovering
the SWM in a geometric way, without invoking the gauge theory
principles. However, the noncommutative feature of the
space-time plays no direct role in the above mentioned
re-derivations of the SWM, with non-commutativity just being postulated in an {\it ad hoc} way.
In the present work, we have shown how
one can construct a noncommutative sector inside an extended phase
space, in a relativistically covariant way. More importantly, we
have shown explicitly how this generalized map can be reduced to
the SWM under certain approximations. Interestingly, this extended
space is physically significant and well studied: It is the space
of the relativistic spinning particle \cite{sg3,sg1}.
 Hence it might be intuitively appealing to think that the NC space-time is
 endowed with spin degrees of freedom, as compared to the ordinary configuration
 space, since the spin variables directly generate the NC {\footnote {This conjecture has been verified by us  in \cite{last}, where we have explicitly constructed a non-commutative target space on a two dimensional manifold. The NC emerges from additional target space spin fields, besides usual space-time coordinate degrees of freedom.}}. The analogue of the
 gauge field is also  identified inside this phase space, without
any need to consider external fields. This situation is to be
contrasted with the NC arising from the background magnetic field
in the well known Landau problem \cite{rev} of a charge moving in a plane in the presence of
a strong, perpendicular magnetic field, or its string theory counterpart \cite{sw,corn}.

We re-emphasize by mentioning that although the coordinate transformation derived here agrees with the previously obtained diffeomorphism \cite{corn}, (as it should), the framework in which it is rederived is entirely distinct from \cite{corn} since here we introduce a dynamical extension of the configuration space intrinsically, whereas \cite{corn} requires an external gauge field. Regarding our identification of the SWM (to $O(\theta )$) as a coordinate transformation  in a specific gauge in the Batalin-Tyutin extension of the spinning particle model, it should be pointed out that the choice of a particular gauge does not restrict the analogy in any way. Because of the gauge invariance of the model, other  gauge choices will simply lead to gauge equivalent theories. In fact one can generate dual systems obeying different gauge conditions which are {\it {not}} noncommutative. Incidentally, this corroborates with the observation of Seiberg and Witten \cite{sw} that the non-c!
ommutative description of a theory is not unique. The above identification, to higher orders in $\theta$, has not been attempted so far but the success in the $O(\theta )$ case is encouraging. Indeed, in itself $O(\theta )$ results are relevant since most of the analysis in NC theories pertain to $O(\theta )$ computations.

 The genesis of the SWM is the observation \cite{sw} that the
non-commutativity in string theory depends on the choice of the regularization
scheme: it appears in {\it e.g.} point-spitting regularization
whereas it does not show up in Pauli Villars  regularization. This
feature, among other things, has prompted Seiberg and Witten
\cite{sw} to suggest the  map connecting the NC gauge fields and
gauge transformation parameter to the ordinary gauge field and
gauge transformation parameter. The explicit form of the
SWM \cite{sw}, for abelian gauge group, to the
first non-trivial order in the NC parameter $\theta _{\mu\nu }$ is
the following,
$$
\hat \lambda (x)=\lambda (x)+{1\over 2}\theta ^{\mu\nu}A_\nu(x)\partial _\mu\lambda (x)
~,$$
\begin{equation}
 \hat A_\mu(x)=A_\mu(x)+{1\over 2}\theta ^{\sigma\nu}A_\nu(x)F_{\sigma\mu}(x)+{1\over
2}\theta ^{\sigma\nu}A_\nu(x)\partial _\sigma A_\mu (x).
 \label{1}
\end{equation}

The above relation (\ref{1}) is an $O(\theta )$  solution of the  general map \cite{sw},
\begin{equation}
\hat A_\mu (A+\delta _\lambda A)=\hat A_\mu (A)+\hat\delta _{\hat\lambda}\hat A_\mu(A),
\label{swm}
\end{equation}
which is based on identifying gauge orbits in NC and ordinary
space-time.

First let us show that it is indeed possible to re-derive the SWM
using geometric objects. We rewrite the SWM (\ref{1})
in the following way,
\begin{equation}
\hat \lambda (x) =\lambda (x)+\frac{1}{2}\{\delta _f[\lambda (x)]-(\lambda
(x')-\lambda (x))\}=\lambda (x)+\delta _f[\lambda (x)]~,
\label{3}
\end{equation}
\begin{equation}
\hat A_\mu (x)=A_\mu (x)+\{\delta
_f[A_\mu (x)]-(A_\mu (x')-A_\mu (x))\}=A_\mu (x)+A'_\mu (x)-A_\mu (x'). \label{4}
\end{equation}
In the above we have defined,
$$
x'_\mu =x_\mu -f_\mu ~~,~~A'_\mu (x')={{\partial x^\nu}\over {\partial
x'^\mu}}A_\nu (x)~~,~~ \lambda '(x')=\lambda (x)~,$$
\begin{equation}
 f^\mu\equiv {1\over
2}\theta ^{\mu\nu}A_\nu ~~. \label{5}
\end{equation}
Here $f^\mu $ is the field dependent space-time translation parameter and
$\delta _f$ constitutes the Lie derivative connected to $f^\mu $,
$$
\delta _f[\lambda (x)]=\lambda '(x)-\lambda (x) =-(\lambda
(x')-\lambda (x))=f^i\partial _i\lambda ,
$$
\begin{equation}
\delta _f[A_\mu (x)]=A'_\mu (x)-A_\mu (x)~.
\label{6}
\end{equation}
This shows that the NC gauge parameter $(\hat \lambda )$ and gauge
field $(\hat A^\mu)$ are derivable from the ordinary one by making
a {\it field dependent} space-time translation $f^\mu$ \cite{j}. One can check
that the NC gauge transformation of $\hat A_\mu (x)$ is correctly reproduced
by considering,
\begin{equation}
\hat\delta \hat A_\mu (x)=\delta (A_\mu(x)+{1\over 2}\theta
^{\sigma\nu}A_\nu(x)F_{\sigma\mu}(x)+{1\over
2}\theta ^{\sigma\nu}A_\nu(x)\partial _\sigma A_\mu (x))~,
 \label{7}
\end{equation}
where $\delta A_\mu(x)=\partial _\mu \lambda (x)$ is the gauge transformation
in ordinary space-time. Hence, if expressed in the form (\ref{4}), the SWM,
(at least to $O(\theta )$), can be derived in a geometrical way, without
introducing the original identification (\ref{swm}) obtained from the
viewpoint of a matching between NC and ordinary gauge invariant sectors. Also note that the
gauge field $A_\mu (x)$ is treated here just as an ordinary vector field, without invoking any
gauge theory properties. This constitutes the first part of our result.

Returning to our starting premises, are we justified in making an
identification between $\hat x_\mu $ in (\ref{nc})-(\ref{2}) and
$x'_\mu $ introduced in (\ref{3})-(\ref{5}), because this relation can connect NC and ordinary
space-time. Naively, a relation
of the form, $x'_\mu =x_\mu -f_\mu (x)$ can not render the
$x'$-space noncommutative, since the right hand side of the
equation apparently comprises of commuting objects only. In our subsequent discussion we will
show how this surmise can be made meaningful and will return to this point at the end.

We start by considering a larger space having inherent NC. Such a space, which at the
same time is physically appealing, is that of the Nambu-Goto model of
relativistic spinning particle  \cite{sg1,sg3}.
 Here the situation is somewhat akin to
the open string boundary NC such that the role of Neveu-Schwarz field is played by here by
the spin degrees of freedom. The Lagrangian of the  model in
2+1-dimensions \cite{sg1,sg3} is,
\begin{equation}
L=[M^2u^\mu u_\mu+{{J^2}\over 2}\sigma ^{\mu\nu }\sigma _{\mu\nu }+MJ\epsilon
^{\mu\nu\lambda}u_\mu \sigma _{\nu\lambda } ]^{{1\over 2}} ~,
\label{8}
\end{equation}
\begin{equation}
u^\mu={{dx^\mu }\over {d\tau }} ~~,~~\sigma ^{\mu\nu }=
\Lambda _\rho ^{~\mu } {{d\Lambda ^{\rho\nu }}\over {d\tau
}}=-\sigma ^{\nu\mu }~~,~~\Lambda _\rho ^{~\mu }\Lambda^{\rho\nu } =\Lambda _{~\rho
}^\mu \Lambda^{\nu \rho }=g^{\mu\nu}~~,~~g^{00}=-g^{ii}=1.
\label{88}
\end{equation}
Here $(x^\mu ~,~\Lambda ^{\mu\nu })$ is a Poincare group element
and also a set of dynamical variables of the theory.

In a Hamiltonian formulation, the conjugate momenta are,
\begin{equation}
P^\mu={{\partial L}\over {\partial u_\mu}} =L^{-1}[M^2u^\mu
+{{MJ}\over 2}\epsilon ^{\mu\nu\lambda}\sigma _{\nu\lambda }]~~,~~
S^{\mu\nu}={{\partial L}\over {\partial \sigma _{\mu\nu
}}}={{L^{-1}}\over 2}[J^2\sigma ^{\mu\nu } +{MJ}\epsilon
^{\mu\nu\lambda}u_\lambda ].
 \label{10}
\end{equation}
The Poisson algebra of the above phase space degrees of freedom are,
\begin{equation}
\{P^\mu ,x^\nu \}=g^{\mu\nu}~~,~~\{P^\mu ,P^\nu \}=0 ~~,~~\{x^\mu
,x^\nu \}=0~~,~~\{\Lambda^{0\mu},\Lambda ^{0\nu}\}=0~,
 \label{011}
\end{equation}
\begin{equation}
\{S^{\mu\nu},S^{\lambda\sigma}\}=S^{\mu\lambda}g^{\nu\sigma}-S^{\mu\sigma}
g^{\nu\lambda}+S^{\nu\sigma}g^{\mu\lambda}-S^{\nu\lambda}g^{\mu\sigma}~,~
\{\Lambda
^{0\mu},S^{\nu\sigma}\}=\Lambda ^{0\nu}g^{\mu\sigma}-\Lambda
^{0\sigma}g^{\mu\nu}~.
\label{012}
\end{equation}
The full set of constraints are,
\begin{equation}
\Psi _1\equiv P^\mu P_\mu -M^2\approx 0 ~~,~~\Psi _2\equiv S^{\mu\nu} S_{\mu\nu}-2J^2\approx 0,
 \label{11}
\end{equation}
\begin{equation}
\Theta _1^\mu \equiv S^{\mu\nu}P_\nu~~~,~~~\Theta _2 ^\mu \equiv \Lambda
^{0\mu}-{{P^\mu }\over M}~~,~~\mu =0,~1,~2~,
 \label{12}
\end{equation}
out of which $\Psi _1$ and $\Psi _2$ give the mass and spin of the particle respectively.
\footnote {Note that instead of $\Psi _2$ as above, one can equivalently use
$\Psi _2\equiv \epsilon
^{\mu\nu\lambda}S_{\mu\nu}P_\lambda -MJ, $
which incidentally defines the Pauli Lubanski scalar.}
 In the Dirac constraint analysis \cite{d}, these are termed as First Class Constraints
 (FCC), having the property that they commute with {\it all} the constraints on the
 constraint surface and
 generate gauge transformations. The set $\Theta _2 ^\mu$ is put by hand \cite{sg3}, to
 restrict
 the number of angular co-ordinates.

The non-commuting set of constraints $\Theta _\alpha ^\mu~,\alpha =1,2 $,
termed as Second Class Constraints (SCC) \cite{d}, modify the
Poisson Brackets (\ref{011}) to Dirac Brackets \cite{d}, defined
below for any two generic variables $A$ and $B$,
\begin{equation}
\{A,B\}_{DB}=\{A,B\}-\{A,\Theta _\alpha ^\mu\}
\Delta _{\mu\nu}^{\alpha \beta }\{\Theta _\beta ^\nu, B\} ,
\label{22}
\end{equation}
\begin{equation}
\{\Theta ^\mu _\alpha  ,\Theta ^\nu _\beta \}\equiv\Delta ^{\mu\nu}_
{\alpha\beta}~~,~~\alpha ,\beta =1,2~~,
~~\Delta ^{\mu\nu}_{\alpha\beta}\Delta _{\nu\lambda}^
{\beta\gamma }=\delta _\alpha ^\gamma\delta ^\mu_\lambda~.
\label{014}
\end{equation}
$\Delta ^{\mu\nu}_
{\alpha\beta}$ is non-vanishing even on the constraint surface.
The main result, relevant to us, is the
following Dirac Bracket \cite{sg1,sg3},
\begin{equation}
\{x_\mu ,x_\nu \}_{DB}=-\frac{S_{\mu\nu}}{M^2} \rightarrow \{\hat x_\mu ,\hat x_\nu \}=
\theta _{\mu\nu}.
\label{db}
\end{equation}
This is the non-commutativity that occurs naturally
in the spinning particle model. Our aim is to express this NC
co-ordinate $\hat x_\mu$ in the form $\hat x_\mu= x_\mu -f_\mu$, with the identification
between $\theta _{\mu\nu}$ and
$S_{\mu\nu}$. This is indicated in the last equality in (\ref{db}).
In the quantum theory, this will lead to the NC space-time
(\ref{nc}).

This motivates us to the Batalin-Tyutin quantization \cite{bt} of the spinning
particle \cite{sg3}. For a system of irreducible SCCs, in this formalism \cite{bt}, the phase
space is extended by introducing additional BT variables,
$\phi
 ^\alpha _a $, obeying
\begin{equation}
\{\phi ^\alpha _\mu,\phi ^\beta _\nu\}=\omega ^{\alpha \beta}_{\mu\nu}=
-\omega ^{\beta \alpha}_{\nu\mu}~~,~~\omega ^{\alpha \beta}_{\mu\nu}=
g_{\mu\nu}\epsilon^{\alpha\beta}~,~\epsilon^{01}=1.
\label{bt}
\end{equation}
where the last expression is a simple choice for $\omega ^{\alpha \beta}_{\mu\nu}$.
The SCCs $\Theta ^\mu_\alpha$ are modified to $\tilde\Theta ^\mu_\alpha$ such that they
become FCC,
\begin{equation}
\{\tilde\Theta ^\mu_\alpha (q,\phi ) ,\tilde\Theta ^\nu_\beta (q,\phi )\}=0
~~;~~\tilde\Theta ^\mu_\alpha (q,\phi )=\Theta ^\mu_\alpha (q)+
\Sigma _{n=1}^\infty \tilde\Theta ^{\mu(n)} _\alpha (q,\phi )~~;~~
\tilde\Theta ^{\mu(n)}\approx O(\phi ^n),
\label{b1}
\end{equation}
with $q$ denoting the original degrees of freedom. Let us introduce the gauge invariant
variables $\tilde f(q)$ \cite{bt}
 corresponding to each $f(q)$, so that $\{\tilde f(q),\tilde \Theta _\alpha ^\mu \}=0$
\begin{equation}
\tilde f(q,\phi )\equiv f(\tilde q)
=f(q)+\Sigma _{n=1}^\infty \tilde f(q,\phi )^{(n)},
\label{b7}
\end{equation}
which further satisfy \cite{bt},
\begin{equation}
\{q_1 ,q_2\}_{DB}=q_3~\rightarrow
\{\tilde q_1 ,\tilde q_2\}=\tilde q_3~~,~~\tilde 0=0.
\label{til}
\end{equation}
It is now clear that our target is to obtain $\tilde x_\mu$ for $x_\mu$. Explicit
expressions for $\tilde\Theta ^{\mu(n)}$ and $\tilde f^{(n)}$ are derived in \cite{bt}.

Before we plunge into the BT analysis, the reducibility of the SCCs
$\Theta ^\mu_1 ~(i.e. ~P_\mu \Theta ^\mu_1$=0)  \cite{sg1,sg3} is to removed \cite{bn}
by introducing a canonical pair of auxiliary
variables $\phi$ and $\pi$ that satisfy $\{\phi ,\pi \}=1$ and PB commute
with the rest of the physical variables. The modified SCCs that appear in the subsequent
BT analysis are as shown below:
\begin{equation}
\Theta _1^\mu\equiv S^{\mu\nu}P_\nu +k_1P^\mu \pi ~~~;
~~~\Theta _2^\mu
\equiv (\Lambda ^{0\mu}-\frac {P^\mu}{M})+k_2 (\Lambda
^{0\mu}+\frac {P^\mu}{M})\phi ~, \label{150}
\end{equation}
where $k_1$ and $k_2$ denote two arbitrary parameters. Since the computations are
exhaustively done in \cite{sg3} they are not repeated here. The results are the following:
\begin{equation}
\tilde x_\mu =x_\mu +
[S_{\nu\mu}+2k_1\pi g_{\nu\mu}](\phi ^1)^\nu
+{{\cal R}_1}_{\mu\nu}(\phi ^2)^\nu +higher-\phi -terms~,
\label{x}
\end{equation}
\begin{equation}
\{\tilde x_\mu ,\tilde x_\nu \}=-\frac{\tilde S_{\mu\nu}}{M^2}~~,~~
\tilde S_{\mu\nu} =S_{\mu\nu}+{{\cal R}_2}_{(\alpha )\mu\nu\lambda}\phi ^{(\alpha)\lambda }
+higher-\phi -terms~,
\label{s}
\end{equation}
where the expressions for ${\cal R}$ are straightforward to obtain
\cite{sg3} but are not needed in the present order of analysis. Only it should be remembered
that the ${\cal R}_1$-term in (\ref{x}) is responsible for the $(\phi ^\alpha )^\mu$-free
term $-S_{\mu\nu}/(M^2)$ in the $\{\tilde x_\mu ,\tilde x_\nu \}$ bracket in (\ref{s}).
Thus the
problem that we had set out to solve has been addressed
successfully in (\ref{x}), which expresses the NC $\tilde x_\mu $
in terms of ordinary $x_\mu $ and other variables \cite{sg3}. 

Now comes the crucial part of
identification of the present map with the SWM \cite{sw}. This means in particular that we have to connect (\ref{x}) to (\ref{5}), since as we have shown before, (\ref{5}) is capable of generating the SWM \cite{sw}. We 
exploit the freedom of choosing  gauges according to our
convenience, since in the BT extended space $\tilde \Theta _\alpha
^\mu $ are FCCs.  For instance, the so called unitary gauge, $\phi
_1^\mu=0,\phi _2^\mu=0$, trivially converts the system back to its
original form  before the BT extension. Let us choose the
following non-trivial gauge,
\begin{equation}
\phi _1^\mu=\frac{M^2}{2}A^\mu (x)~~,~~\phi _2^\mu=0,
\label{gauge}
\end{equation}
where $A^\mu (x)$ is some function of $x_\mu$, to be identified with the gauge field.
Let us also work with terms linear in $A^\mu (x)$. Identifying $\tilde S_{\mu\nu}/(M)^2=
\theta _{\mu\nu}$ we end up with the cherished mapping,
\begin{equation}
\tilde x_\mu =x_\mu -\frac{1}{2}\theta _{\mu\nu}A^\nu (x) +higher-A(x)-terms ~,
\label{b115}
\end{equation}
\begin{equation}
\{\tilde x_\mu ,\tilde x_\nu \}_{DB}
=\theta _{\mu\nu} +higher-A(x)-terms~ .
\label{db3}
\end{equation}
Note that in the above relations (\ref{b115},\ref{db3}), we have dropped the terms
containing $k_1$, an arbitrary parameter \cite{sg3}, considering it to be very small.
Also in (\ref{db3}) Dirac Bracket reappears since the system is gauge fixed and hence has SCCs. This constitutes the second part of our result.

Finally, two points are to be noted. Firstly, the
non-commutativity present here does {\it not} break Lorentz
invariance since there appear no constant parameter
with non-trivial Lorentz index to start with. The violation will
appear only in the identification of $\tilde S_{\mu\nu}$ with
(constant) $\theta _{\mu\nu}$. Secondly, (\ref{x}) truly expresses
the NC space-time $\tilde x_\mu$ in terms of ordinary space-time
$x_\mu$. But $x_\mu$ becomes NC owing to the Dirac brackets
induced by the particular gauge that we fixed in order to reduce
our results to the SWM. Obviously, in general, there is no need to
fix this particular gauge. This refers to the comment below (\ref{7}).

To conclude, we have shown that it is possible to view the
(abelian $O(\theta )$) Seiberg Witten map as a co-ordinate transformation involving
field dependent parameters. The idea of equivalence between gauge orbits in noncommutative and
ordinary space-times, which was crucial in the original derivation \cite{sw}, is not applied
here.
It has been explicitly demonstrated that a noncommutative space-time sector can be constructed
in the Batalin-Tyutin extension of the relativistic spinning particle model \cite{sg3}.
Finally, the above mentioned transformation and subsequently a direct connection with the
Seiberg-Witten map is also generated in this model. It emerges from the present work that
noncommutative space-time is endowed with spin degrees of freedom, as compared to the
ordinary space-time \cite{last}.

\vskip 1cm Acknowledgement: It is a pleasure to thank Professor
R.Jackiw for helpful correspondence.

\end{document}